\documentstyle[aps,manuscript]{revtex}

\begin{document}
\draft
\title{Effects of Columnar Pins on Flux Line Dynamics}
\author{Seungoh Ryu, A. Kapitulnik and S. Doniach }
\address{Department of Applied Physics, Stanford University, Stanford,
CA94305}
\maketitle

\begin{abstract}
The effects of columnar pins on the flux lines in a model High-Tc
superconductor in an applied field are studied through Monte Carlo
simulations. An analytic solution for a single line
case is obtained and compared with  the simulation results. By introducing
a tilted potential,  we study simulated ``IV" characteristics and the
results  indicate a distinct scaling behavior above and below the
depinning  temperature. We introduce a diverging length scale  measured in
terms of a ``retrapping length'' to analyze self-similar behavior across
the depinning  transition.
\end{abstract}


The nature of the mixed state of High-$ {\rm T_c} $ and of artificially
layered
superconductors has been the subject of intense debate during the past
few years\cite{nelson88,fisher91}.
In these layered superconductors, vortices may be described as two
dimensional disks with weak interlayer coupling hence allowing the vortex
to fluctuate in all directions even for ${\rm T \ll
T_c}$\cite{lawrence70,bulaevskii91,clem91}.  It is now
believed\cite{houghton89,doniach89,ryu92,safar92} that in  clean
materials, the flux lattice can melt at temperatures significantly  below
${\rm T_c }$ and the notion of a melted flux liquid is relevant  for a
large region of the H-T phase diagram. A clear understanding of the
kinetics of the viscous flux liquid in the presence of the pinnning
potential is very  important, especially in efforts to control dissipation
induced by the  moving flux lines. Recently, proton and heavy ion
irradiation have been  used to introduce well controlled defects into the
material to enhance the
pinning\cite{civale90,lombardo92,civale91,gerhauser92,worthington92}. Both
techniques turned out to increase the critical current density
significantly while moving the ``irreversibility line'' to higher field
and  temperature region for both YBCO and BSCCO single crystals.  In this
letter, following earlier work of Nelson and  Vinokur\cite{nelson92}, we
show analytically and numerically that the temperature induced depinning
of a single flux line in the presence of a columnar pin has a {\it quasi}
critical point ${\rm T_{dp}}$ with scaling behavior that dominates the
kinetics close to ${\rm T_{dp}}.$

There is an elegant way to describe the system by mapping the
three dimensional flux lines onto the world-lines of a two-dimensional
boson system\cite{nelson92}.
We consider a long flux line in a system of ${N_z= L/d} $ layers (indexed
z). {\it L} is the total thickness of the sample and {\it d} is the
distance between the superconducting layers.
The position of the pancake vortex\cite{clem91} in the {\it z}-th plane is
given by $\{ {\bf r}_{z} \}.$
The flux line interacts with a columnar pin of radius {\it a} placed at
the origin parallel to the z-axis.
Pancakes in the adjacent layers are coupled
through the interlayer interaction for which we assume a quadratic form for

displacements smaller than $r_J \equiv 2 \xi_{ab}/\sqrt{g} $ and a
linear form for larger separations\cite{doniach89,feigelman90}. $ \sqrt{g}
=  {\xi_{c} \over d} $ is the small interlayer coupling
parameter of the Lawrence-Doniach model\cite{lawrence70}.
A detailed description for more general cases is given in \cite{ryu92}.

The normalized partition
function for this system is then given by
%
%
\begin{eqnarray}
\label{singleaction}
{Z}_{N_z} & = & \prod_{z=1}^{N_z} \int \! d{\bf r}_z \,e^{ - \beta
\sum_{z=1}^{N_z} (
\epsilon |{\vec
r}_{z}-{\vec r}_{z+1}|^2 - \eta U \Theta( a-|{\vec r}_z| )) }
/ ({ \pi \over \beta \epsilon })^{N_z}
\nonumber \\ & = & \int \! d{\bf r}_1 \! \int \! d{\bf r}_2 \;
\zeta_{N_z}({\bf r}_1,{\bf r}_2) / ({ \pi \over \beta \epsilon })^{N_z}  \,
, \end{eqnarray}
where in the second line, the partition function is expressed as a sum
over configurations with both ends fixed at $ ({\bf r}_1,{\bf r}_2).\; $
$\beta $ is the inverse temperature, ${1 \over k_BT}$ and $\epsilon = c
\cdot {\rm U / r_J}^2 \, $ where c is a constant of order unity and ${\rm
U} = {{\rm d} \phi_0^2 \over 8 \pi^2 \lambda^2 } .\ $
$\phi_0 $ is the flux quantum and $\lambda = \lambda_0 / \sqrt{1- {\rm
T/T_{c0}}} \ $ is the in-plane penetration depth. Finally, the last term in

the exponent is due to the pinning potential which is represented by a
cylindrical potential well of radius {\it a} along z-axis with uniform
depth given as a fraction $\eta $ of U. $\zeta_{N_z}({\bf r}_1,{\bf r}) $
for an $N_z $ layered system satisfies the following iterative relationship

%
\begin{eqnarray}
\zeta_{N_z\!+\!1}({\bf r}_1,{\bf r})=e^{\beta \eta U
\Theta(a\! -\! r)} \! \int \! d{\bf r'}\, e^{- \beta
\epsilon |{\bf r} -{\bf r'}|^2 } \! \zeta_{N_z}({\bf
r}_1,{\bf r'}) / ({\pi \over \beta \epsilon }).
\nonumber
\end{eqnarray}
We may write the solution in the form,\cite{lifshitz69,degennes69}
%
%
\begin{eqnarray}
\label{partition}
\zeta_{N_z}({\bf r}_1,{\bf r}) = \sum_{i}
e^{-N_z\, \lambda_i}
f_i^*({\bf r}_1)f_i({\bf r}),
\end{eqnarray}
where $\{ f_i({\bf r}) \} $ are the eigenfunctions of the following 2D
equation with eigenvalues $\lambda_i,$

%
%
\begin{eqnarray}
\label{eigenvalue}
{1 \over 4 \beta \epsilon}\,\nabla_{xy}^2 \, f_i({\bf r}) + f_i({\bf r}) =
e^{-\beta \eta U \Theta(a\! -\! r)-\lambda_i} f_i({\bf r})\, .
\end{eqnarray}
By expanding the exponential term, this problem can be converted into a
simple Schr\"odinger equation for a particle in a cylindrical
well\cite{nelson92}. However, as we show later, $-\beta \eta U +|\lambda_i|

> 1 $ for all temperatures of interest and it is desirable to solve
Eq.~\ref{eigenvalue} as it is. Due to the simple form of our potential
well, we can still turn it into a Schr\"odinger form by redefining
$e^{-\lambda_i} \equiv -E.$ Then $E < 0 $ for real values of $\lambda . $
Note that the equation then describes a particle with different values of
mass inside:m=$2\beta \epsilon \hbar^2 e^{-\beta \eta U} $ and
outside the well:m=$2\beta \epsilon \hbar^2 $ with depth of the well given
by $|V| = e^{\beta \eta U }-1. $ We look for bound states $(E < -1) $
satisfying the condition:
${\rm J_n}(Ka) / (e^{\beta \eta U} K \cdot {\rm J'_n}(Ka))= {\rm
H_n}^{(1)}(ika) / ik \cdot {\rm H_n'}^{(1)}(ika), $
where ${\rm J_n, \/ H_n}^{(1)} \/ $ are the standard Bessel functions and
$K = \sqrt  {4\beta\epsilon (1 -|E|\, e^{-\beta \eta U})}, \/ k = \sqrt
{4\beta \epsilon(1-|E|)}.$\
This condition is derived from noting that the statistical
weight of finding the particle at ${\bf r}  $ in the z-th layer is given
by

%
%
\begin{eqnarray}
\label{density}
{\rm n_z}({\bf r})\propto  \int d{\bf r}_1 \int d{\bf r'} \; \zeta_{z}({\bf

r}_1,{\bf r}) \; \zeta_{N_z - z}({\bf r},{\bf r'}),\
\end{eqnarray}
and requiring that the particle flux as well as the density be continuous
across the boundary of the well.

Let us briefly describe how our Monte Carlo simulation is
implemented for  this model system. The xy-coordinates of each pancake
vortex in the layer  are taken as physical variables. The flux line is
placed in a box of $222  \times 256 \times 1024$ grid space with periodic
boundary conditions in  the xy-plane\cite{ryu92}. The columnar pin is
modeled as a cylindrical potential well placed at the center of the box and
periodically repeated through the boundary conditions.
For the initial configuration, we use a
straight flux line placed  near(within 30 grids spacing) the potential
well at the origin. To achieve  equilibrium for each temperature, at least
52,000 MC steps using Metropolis algorithm were followed while measuring
the equilibrium properties(Figures 1, 2).  We then turn on a uniformly
tilted potential to simulate an applied transport current density in an
``IV" type measurement(Figures 3, 4). Following  at least 1000 MC-steps
using a Kawasaki algorithm \cite{Bortz75} after switching on the tilted
potential to attain  steady state, we measure the average distance over
which the center of mass$(X_{cm})$ of the flux line moves for a duration of
$\tau$(5000- 50000  MC-steps)\cite{footnote1}. We relate the measured
drift velocity, $(X_{cm}(\tau )-X_{cm}(0))/\tau$, to the voltage through
the Josephson relation\cite{tinkham}.

In Figure~\ref{singleline}, we compare the results of a single line
simulation(1024 layers) to the analytic evaluation of the pinned fraction,
${\rm f_p}, $ as the temperature is varied. The fraction of
the pinned pancake  vortices in equilibrium, ${\rm f_{p}},$ is evaluated
from ${1 \over N_z} \sum_{z=1}^{N_z} \int  d{\bf r} \langle \Theta (a-|\bf
{r}| ) \, {\rm n_z}(\bf {r}) \rangle $ in our simulation where $ \langle
...\rangle $ means
the thermal average and $\Theta (x) $  is the unit step function.
${\rm n_z}(\bf {r})$ is as defined in Eq.~\ref{density}.
In the
analytic calculation we take the  thermodynamic limit$(N_z \rightarrow
\infty)$ and the lowest energy bound  state, $g_{\lambda_g}({\bf r}), $
dominates the sum in Eq.~\ref{partition} so that ${\rm f_p }= \int_0^a
|g_{\lambda_g}(r)|^2\  r dr / \int_0^\infty  |g_{\lambda_g}(r)|^2\  r dr .
$
The calculation was run with two different column radii, ${a \over
\lambda_{ab}} = 0.01$(not shown) and $0.05$~.
We obtained an excellent agreement between the analytic result(solid
line) and the simulation(filled points) without any adjustable
parameters.
When the radius of the pin is increased five-fold, the curve shifts
toward higher temperatures by about 10 \% . The inset shows the radial
distributions of the pancakes, ${\rm n(r)} = r |g_{\lambda_g}(r)|^2, $
across the transition temperature.
The distribution has a peak inside the column in the deeply bound region
while for higher temperatures, the peak
gradually spreads out of the column and the distribution is virtually
uniform. It is  well known from elementary quantum mechanics that the 2D
Schr\"odinger  equation with a well potential has always a bound state
however weak the  potential. Therefore, we can not associate our
observation with a true  phase transition associated with dissappearance
of the bound state \cite{footnote2}, however, operationally we can define
(within  some termperature interval $\triangle {\rm T} = \pm 1 {\rm K} $)
 a depinning transition
temperature$({\rm T_{dp}})$ beyond which the vortex distribution is
unifrom across the finite system. Fluctuations in the internal
energy show a distinct specific heat peak in this region and we use
this to define ${\rm T_{dp}}$ for which we find that ${\rm f_p} \approx
0.05. $ In finite fields B for which the
mean vortex spacing is less than the average pin spacing and there are
several meandering flux lines  competing to occupy a given columnar
pin, lines will have a finite probability per unit length of
being displaced by other flux lines. This can be represented by a reduced
effective column length which depends on B. This will define a depinning
boundary  ${\rm B_{dp}(T)}$ in the B-T phase diagram.

As ${\rm T_{dp}} \/ $ is approached from below, thermal fluctuations in the

form of freed segments of various lengths develop. We find it useful to
identify a retrapping length, $\xi_{rt} $
defined as the {\it  maximum} length of the continuously freed portion of
the line
measured  along the z-axis.
An operator to project out a continuously freed segment of size {\it l}
in units of {\it d} can be defined as $\hat{l} = \sum_{i=1}^{N_z-l}\hat{\rm
P}_{i-1} \hat {\rm P}_{i+l} \prod_{j=1}^{l}(1-\hat{\rm P}_{i+j-1}) $
where $\hat{\rm P}_i $ gives 1 if the particle in the {\it i}-th layer is
inside the well and 0 otherwise. Using the projection
operator defined above, the
probability distribution for segments of length {\it l} as a function of
temperature, ${\cal P}({\it l}, {\rm T}), \/ $ can be derived as:

\begin{equation}
	{\cal P}({\it l}, {\rm T})\sim e^{-|\lambda_g | l}\int_0^\infty
d\varepsilon
	{\cal D}(\varepsilon ) e^{-l \varepsilon } |\langle
g_{\lambda_g}|g'_\varepsilon
	\rangle |^2
	\label{distribution}
\end{equation}
where $\lambda_g, |g_{\lambda_g}\rangle $ are the eigenvalue and
eqigenstate of the ground
state from Eq.~\ref{eigenvalue} and $|g'_\varepsilon \rangle $ and ${\cal
D}(\varepsilon ) $ are the continum states and their density calculated
with a similar equation with a hard core repulsive potential replacing the
attractive pinning well. The hard core potential over a distance of {\it l}

is needed to keep the flux line out of the well and thus represents the
projection operator introduced above in defining the free segment of size
{\it l}. $ | \langle g_{\lambda_g}|g'_\varepsilon \rangle |^2 \/ $ is the
overlap integral between the two eigenfunctions.
As ${\rm T_{dp}} $ is  approached from below, the binding energy$(|
\lambda_g |)$ as determined by numerical solution of Eq.~\ref{eigenvalue}
approaches zero as $\sim (1 - {\rm {T \over T_{dp} } })^{\nu} \equiv
t^{\nu }$  with $\nu \approx 1.1\pm 0.2. $ The error is a measure of
the sensitivity of the exponent to an uncertainty of $\pm 1$ K
in the definition of ${\rm T_{dp}} $.
The curves in Fig.\ref{xidistrib} show the probability distribution as
measured in our simulation. The series of curves for different
temperatures  show a striking self-similar behavior which can be expected
from the power law factor in the formula derived above.
 From our simulation, we find that $\xi_{rt} \approx  t^{-\nu} $ for ${\rm
T < T_{dp}.}$ This is represented in the inset of  Fig.~\ref{xidistrib}
where $\xi_{rt} $ and ${1 \over | \lambda_g |} $ are  plotted versus {\it
t}. Note that this {\it approximate} power law behavior is observed {\it
over a wide range  of temperature} $(0.7 {\rm T_{dp}} \le {\rm T} \le {\rm
T_{dp}}).$ This  almost critical behavior provides a convenient concept to
describe the  scaling of the ``IV" curves in our simulation.
Nelson and Vinokur have introduced a localization length
scale$(l_{\perp})$ for a similar system in which discrete  layers are
treated in the continum approximation\cite{nelson93}. Since a free segment
of size {\it l} will on average extend over a lateral area of $l/2\beta
\epsilon$ in the ab-plane, we can relate $<l>$ to $l_{\perp}$ through  $2
l_{\perp}^2 = \int dl\, {\cal P}(l,{\rm T}) l/\beta \epsilon ={ <l>/\beta
\epsilon}.$
In the situation where $l_{\perp}$ spans over
more than one columnar pins, $l_{\perp}$ will be the more natural length
scale to characterize the system\cite{balents93}.
However, $\xi_{rt}$ which we relate
directly to $1/|\lambda_{g}|$ is more useful for
discussing the kinetics.

To study how the approximate self-similarity above manifests itself in
the non-equilibrium situation, we consider the kinetics of a long line
drifting along the uniformly tilted potential with a periodic array of
columnar pins.
Figure~\ref{reptation} shows a typical snapshot of the final
configurations at 5000 MC steps after switching on the tilt potential
equivalent to the current density of $6.67  \times 10^7 {\rm A/cm^2}.$ The
periodic vertical lines represent  the positions of the columnar pins.
At temperatures close to and above ${\rm T_{dp}}, $  the lines drift as a
whole with velocity linear to  the applied current. For T lower than ${\rm
T_{dp}}, $ the  pinning is relatively strong and the line moves along in
the tilt direction  by stretching out a ``free'' finger to the nearby
pinning sites(low current  case).
Since these processes involve
formation and motion of free segments of average  diameter $\xi_{\rm
rt}$, they will introduce a time scale $1 /\omega \approx  \xi_{\rm
rt}^{z'} $ with a dynamical exponent $z'$.

The results of the ``IV'' measurements in our
simulation (inset to  Figure~\ref{singleiv}) show a crossover from
non-linear to linear behavior  across the depinning temperature. Using the
controlling length scale of  $\xi_{\rm rt} $ and a dynamic scaling law $1
/ \omega \propto \xi_{\rm rt}^{z'}\  $ for the dominant relaxation
mode\cite{fisher91}, we can collapse the curves to a  single scaling
function with two distinct branches of the form:
\begin{equation}
\label{scaling}
V = \,\xi^{-(1+z')}_{rt} F_\pm ({\rm J \over T}\xi^{1+z}_{rt})
\end{equation}
with $\nu  (1+z) = 2.75\pm 0.2, \nu (1+z') = 3\pm 0.2. $
The resulting scaling behavior is somewhat similar
to the results of Lee {\it et al.}~\cite{lee93}.
They studied a case where the inter-vortex
interaction is important. Thus our results on a single line seem to suggest
that the scaling behavior  observed in both cases is generic to a flux
line interacting with correlated pins.
We also find that the
experimental data of Worthington {\it  et al.}\cite{worthington92}
can be well fit by Eq.~\ref{scaling} in the high current regime.
Thus, interpretations of
nonlinear ``I-V" characteristics in the relatively high current
density region with correlated strong pins may be interpretable in terms
of the single line limit. It is also an interesting and
open question how far the scaling behavior observed in our system with
columnar pins can be extended to the case of random point  pins in a dense
flux liquid regime where it may be expected that vortices  will
seek out a maximally pinned path. It is then possible that the range of
validity of the
columnar model may extend to the more general case.

We thank Dr.~M.~Suenaga and John R.~Clem for stimulating discussions. AK
thanks David Nelson for fruitful discussions. SR thanks Leonid Pryadko
for the Friday night discussions. Supports by AFOSR Grant No.~91-0145 by
the Center for Materials Research through the MRL/NSF program and
RP8009-24 by EPRI are gratefully acknowledged.


%
%

\newpage
\baselineskip = 2\baselineskip

\begin{figure}
\caption{ Pinned fraction of a single flux line in 1024 layer system (Dots:
MC  simulation, Solid line:analytic evaluation) for a system with $a
/ \lambda_{ab} = 0.05.$ In the region where dots from the simulation
deviate from the analytic results, the flux line wanders beyond the unit
bounding box of our simulation and due to the periodic boundary
conditions,  begins to probe the square array of columns which are
separated by $12.8 \lambda_{ab}$. It is clear that the ``delocalization"
condition as discussed in \protect\cite{nelson92}
 is met in this region.
The inset shows the radial distribution functions, ${\rm n(r)} =
r |g_{ \lambda_g }(r)|^2$ .}\label{singleline}   \end{figure}

\begin{figure}
\caption{ Probability distribution of the free segment of size {\it l}
when the tilt is zero. ${\rm T / T_{dp}}$ for each curve runs from
0.47 to 1.3 where the curve with the largest extent belongs to the highest
temperature. The  inset shows the dependence of $\xi_{rt} $ and
$1/\lambda_g$ on t$=1 - {\rm
T/T_{dp}} $ (Dots :MC simulation, Solid curve:analytic evaluation).
}\label{xidistrib}
\end{figure}

\begin{figure}
\caption{ Snapshots taken at t = 5000 MC steps from the ``IV" measurements
for different  values of temperature for the current density of $6.67
\times 10^7 {\rm  A/cm^2}. $ The tilt is along the x-direction and all the
lines initially  started as straight located at x=128. The periodic
vertical lines are the  columnar pins. The numbers attached to the lines
are ${\rm T/T_{dp}} $ values.}\label{reptation}
\end{figure}

\begin{figure}
\caption{ Result of the ``IV" measurements from simulation. The inset shows
the raw  data for ${\rm T/T_{dp}} $ ranging from 0.47 to 1.3. They can be
collapsed to two scaling curves using the following exponents as discussed
in the text: ${\rm T_{dp}/T_{c0}} = 0.74, \nu (1+z) = 2.75, \nu (1 + z') =
3$ .}\label{singleiv}
\end{figure}


\begin{references}
\bibitem{footnote2}
\newblock
 From the solutions to Eq.~3, we find that there  always is at least one
bound state for each temperature except for very low temperatures$({\rm T}
< 30 K)$ where there are more than one. We also observe that $\beta \eta
U-\lambda_g>1$ for ${\rm T>T_{dp}}.$
\end{references}
\end{document}